\documentclass[twocolumn,pre,showpacs,superscriptaddress]{revtex4}
\usepackage{epsfig,amssymb,amsmath}
\usepackage{color}

\begin{document}
\title{Structured Chaos in a Devil's Staircase of the Josephson
  Junction} \author{Yu. M. Shukrinov} \affiliation{BLTP, JINR, Dubna,
  Moscow Region, 141980, Russia} \author{A. E. Botha}
\affiliation{Department of Physics, University of South Africa,
  P.O. Box 392, Pretoria 0003, South Africa} \author{S. Yu. Medvedeva}
\affiliation{BLTP, JINR, Dubna, Moscow Region, 141980, Russia}
\affiliation{Moscow Institute of Physics and Technology (State
  University), Dolgoprudny, Moscow Region, 141700, Russia}
\author{M. R. Kolahchi}  \affiliation{Institute for Advanced Studies
  in Basic Sciences, P.O. Box 45195-1159, Zanjan, Iran}
\author{A. Irie} \affiliation{Department of Electrical and Electronic
  Systems Engineering, Utsunomiya University, 7-1-2 Yoto, Utsunomiya
  321-8585, Japan}

\date{\today}

\begin{abstract}
The phase dynamics of Josephson junctions under external
electromagnetic radiation is studied through numerical
simulations. Current--voltage characteristics, Lyapunov exponents and
Poincar\'{e} sections are analyzed in detail. It is found that the
subharmonic Shapiro steps at certain parameters are separated by
structured chaotic windows. By performing a linear regression on the
linear part of the data, a fractal dimension of $D=0.868$ is obtained,
with an uncertainty of  $\pm 0.012$.  The chaotic regions exhibit
scaling similarity and it is shown that the devil's staircase of the
system can form a backbone that unifies and explains the highly
correlated and structured chaotic behavior. These features suggest a
system possessing multiple complete devil's staircases. The onset of
chaos for subharmonic steps occurs through the Feigenbaum period
doubling scenario. Universality in the sequence of periodic windows is
also demonstrated. Finally the influence of the radiation and
Josephson junction parameters on the structured chaos is investigated
and it is concluded that the structured chaos is a stable formation
over a wide range of parameter values.
\end{abstract}
 \maketitle

\section{INTRODUCTION}
Nonlinear dynamical systems like the Josephson junction exhibit  a
wide variety of chaotic phenomena and are an interesting field of
research for both the fundamental and applied sciences
\cite{likharev86,kleiner04}. Much progress has been made through
numerical simulations of many important characteristics of chaos
\cite{kautz96}. Investigations of critical exponents in various routes
to chaos, such as period-doubling bifurcations, intermittency, and
quasiperiodicity lend support to the universality and scaling behavior
predicted by theoretical models \cite{he85}. In a Josephson junction
driven by an external microwave radiation, the devil's staircase
structure is realized as a consequence of the interplay of the
Josephson plasma frequency, and the applied frequency (see
Refs.\cite{benjacob81,jensen83} and references therein).

The devil's staircase appears in other systems including the infinite
spin chains with long-range interactions \cite{nebendahl13},
frustrated quasi-two-dimensional spin-dimer system in magnetic fields
\cite{tarigawa13},  systems of strongly interacting Rydberg atoms
\cite{weimer10}, and in fractional quantum Hall effect
\cite{laughlin85}. The synchronization of Josephson and Bloch
oscillations results in the quantization of transresistance which also
leads to a devil's staircase structure.\cite{hriscu13} Devil's
staircase in a spin-torque nano-oscillator driven by a microwave field
was experimentally demonstrated in Ref. \cite{urazhdin10}.

In earlier studies of the rf-current-driven JJ it was found that,
depending on the amplitude of the rf current, the system would develop
a chaotic character \cite{huberman80,kautz81,kautz85a,kautz85b}. It
was also shown that the onset of chaos as a function of frequency
correlated with the function that indicated infinite gain, also as a
function of frequency, for the unbiased parametric amplifier
\cite{pedersen81}. More recently the importance of chaos in intrinsic
JJs, and its effects on the IV-characteristics and the Shapiro steps
in these systems, were stressed in Refs.~\cite{irie03,scherbel04}.

In a further comprehensive study, the critical behavior of the
dynamical equation, in the sense of how it goes from regular to
chaotic, was investigated \cite{bohr84}. It was discovered that the
critical behavior is that of the circle map. Here, in the subcritical
state, the resonances are separated by quasi-periodic orbits
(i.e. having irrational winding number), whereas in the supercritical
state the dynamics is composed of chaotic jumps between
resonances. The chaos appears to be the result of the resonance overlap \cite{jensen84}.
In a similar  study, the interaction of the
Josephson junction with its surroundings was considered by coupling it
in parallel with an RLC circuit, modeling a resonant cavity
\cite{borcherds87}. In this case the chaos develops through the
familiar infinite sequence of period doubling bifurcations.

Other studies exist that try to understand the irregular response of
dissipative systems driven by external sources when such behavior is
interleaved by the synchronized motion (Shapiro steps). This so called
intermittent chaos is modeled by a random walk between the two
neighboring steps that have become unstable. A simple physical model
was proposed in Ref.~\cite{ben-jacob82} and used for an analytic
calculation of the power spectrum of the intermittent-type chaos
occurring in rf- and dc-current-driven Josephson junctions. It was
shown that the power spectrum of the voltage correlations has a
broadband background characterizing the chaotic solution. A similar
random walk model was developed to explain the phase synchronization
of chaotic rotators \cite{osipov02}.

The microwave-induced devil's staircase structure and chaotic behavior
in current-fed Josephson junctions were studied experimentally and
through analog simulations within the RSCJ model in
Ref.~\cite{ben-jacob81}.  Many externally driven dissipative systems
demonstrate the occurrence of sequences of periodic states separated
by gaps of a chaotic or intermittent nature.\cite{ben-jacob82}.
Different features and types of intermittent transitions to the
chaotic state were investigated in
Refs.~\cite{yeh84,ben-jacob82,seifert84,seifert82-proceeding,noeldeke85}.
The RCSJ model was used to study the onset of chaotic behavior in the
overdamped JJ, where a transition from unlocked quasiperiodic motion
via subharmonic locked steps to the chaotic region occurs
\cite{seifert83,jensen83}. In dynamical systems, intermittency is the
temporal irregular alternation of phases between apparently periodic
and chaotic dynamics. Pomeau and Manneville described three routes to
intermittency when a nearly periodic system shows irregularly spaced
bursts of chaos  \cite{manneville79,pomeau80,pomeau84}. These, types
I, II and III respectively, correspond to the approach to a
saddle-node bifurcation, a subcritical Hopf bifurcation, and an
inverse period-doubling bifurcation. Type III intermittency was also
observed in Ref.~\cite{noeldeke85}. In all the above mentioned types
of intermittency the width of each phase is unpredictable.

We could also have other physical systems with competing frequencies
leading to the devil's staircase structure. In a system with magnetic
vortex oscillations, the sense of gyration of the magnetic vortex
depends on the vortex core polarity, and can undergo sudden reversals
as a critical velocity is reached.  Driving currents can tune the
self-sustained vortex oscillations and the frequency of the core
reversals, resulting in phase-locked or chaotic states, within a
devil's staircase structure \cite{sebastien12}. In two capacitively
coupled Josephson junctions, the phase-locked structure of the
oscillations forms a devil's staircase, and chaotic dynamics develops
between the main resonances as this coupling capacitance passes a
critical value \cite{valkering00}. Another realizable model that goes
chaotic via the resonance overlap in the devil's staircase structure,
involves a modified Chua's circuit \cite{furui14}. A class of models
involve transport in an external potential; the resonance manifests
itself in an enhanced diffusion constant \cite{reimann02}. Addition of
competing forces, including noise, could result in anomalous transport
and chaos; the predicted absolute negative mobility could be tested in
a resistively and capacitively shunted Josephson junction
\cite{speer07}.  The ensuing devil's staircase in the dynamics of a
charged particle as it moves in electrostatic waves is a rare instance
of a time dependent potential for which the effect has been observed
\cite{macor06}.

In the present paper, we demonstrate a novel type of chaos, which for
good reasons, is called structured chaos. The chaotic behavior we deal
with here is set against the devil's staircase structure in the
IV-characteristics of an underdamped Josephson junction at some
parameters of the system, and in the presence of external radiation.
The positions of the steps in the structure is determined by a
continued fraction formula \cite{S1}, so that the transformations of
the regular behavior in the step's current intervals to the chaotic
ones is strongly synchronized.  We show that in the chaotic windows we
see remnants of this staircase, and we present more details of this
structurally chaotic behavior. We discuss the role of the subharmonic
subset of steps in the staircase, in bringing about such a structure,
and discuss ideas having to do with the presence of multiple
staircases \cite{qu98}.

The paper is organized as follows. In Sec. II we introduce the model
and briefly describe the simulation procedure and parameters of the
model and simulations. The IV-characteristics of the JJ under external
radiation demonstrating structured chaos (alternating changes of the
steps and chaotic regions, with changing bias current) are
presented. We analyze the main features of this structure and
calculate its fractal dimension. The chaotic nature of the
IV-characteristic portions between steps is discussed based on high
precision calculation of the Lyapunov exponents and the Poincar\'{e}
section in Sec. III. In Sec. IV we discuss some unifying properties,
particularly, the backbone features and Farey sum rule. The scaling
features and on-step positive LE are analyzed in Sec. V. In Sec. VI we
present the results of the influence  of the radiation and JJ
parameters on the structured chaos. Sec. VII is devoted to the
analysis of the existing experimental results. Finally, Sec. VIII
concludes the paper.

\section{SUBHARMONIC STEPS}
First we discuss the features of the JJ IV-characteristics under
external electromagnetic radiation. To simulate the IV-characteristics
we use the resistively and capacitively shunted junction (RCSJ) model
\cite{stewart68,mccumber68} for the Josephson junction driven by both
ac and dc current sources.  This model is equivalent to that for a
pendulum with dissipation, and driven by an external torque with both
constant and periodic components. The model equations for the phase
difference $\varphi$ across the junction, taking into account the
external radiation with frequency $\omega$ and amplitude $A$, are
\begin{eqnarray}
\dot{V}+\sin(\varphi)+\beta\dot{\varphi}=I+A\sin(\omega t) \mbox{,}
\label{current} \\ \dot{\varphi}=V \mbox{.}
\end{eqnarray}
Here the dc--bias current $I$ and ac amplitude $A$ are normalized to
the critical current $I_c$,  the voltage $V$ to $V_0=\hbar
\omega_p/(2e)$, where $\omega_p$ is the plasma frequency, and time $t$
to $\omega_p^{-1}$. The dissipation parameter is
$\beta=\beta_c^{-1/2}$, where $\beta_c$ is McCumber's
parameter. Overdot indicates derivative with respect to the
dimensionless time $t$.

To simulate the experimental conditions we have tested the effect of
adding white noise to the bias current, with amplitude $I_{\rm noise}
= 10^{-8}$. The amplitude of the noise current was also normalized to
the critical current $I_c$. We found that the features of
IV-characteristics discussed in this paper do not depend on the noise
in current, at this amplitude.

In the numerical simulations for this study we set  $\beta=0.3$
(mostly), and  $I_{\rm noise}=0$. We make use of a fourth-order
Runge-Kutta integration scheme, using a time step of $1/32$, with
$10^{4}-10^{5}$ as a time domain for averaging, $10^3-10^{5}$ units
before averaging, and $10^{-5}-10^{-6}$ as the step in the bias
current. Further details concerning the simulation procedure can be
found in Refs.~\cite{sg-prb11,pla12}.

\begin{figure}[htp!]
\includegraphics[height=65mm]{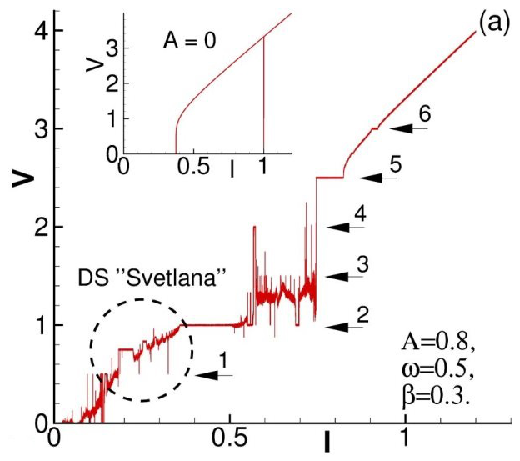}
\includegraphics[height=65mm]{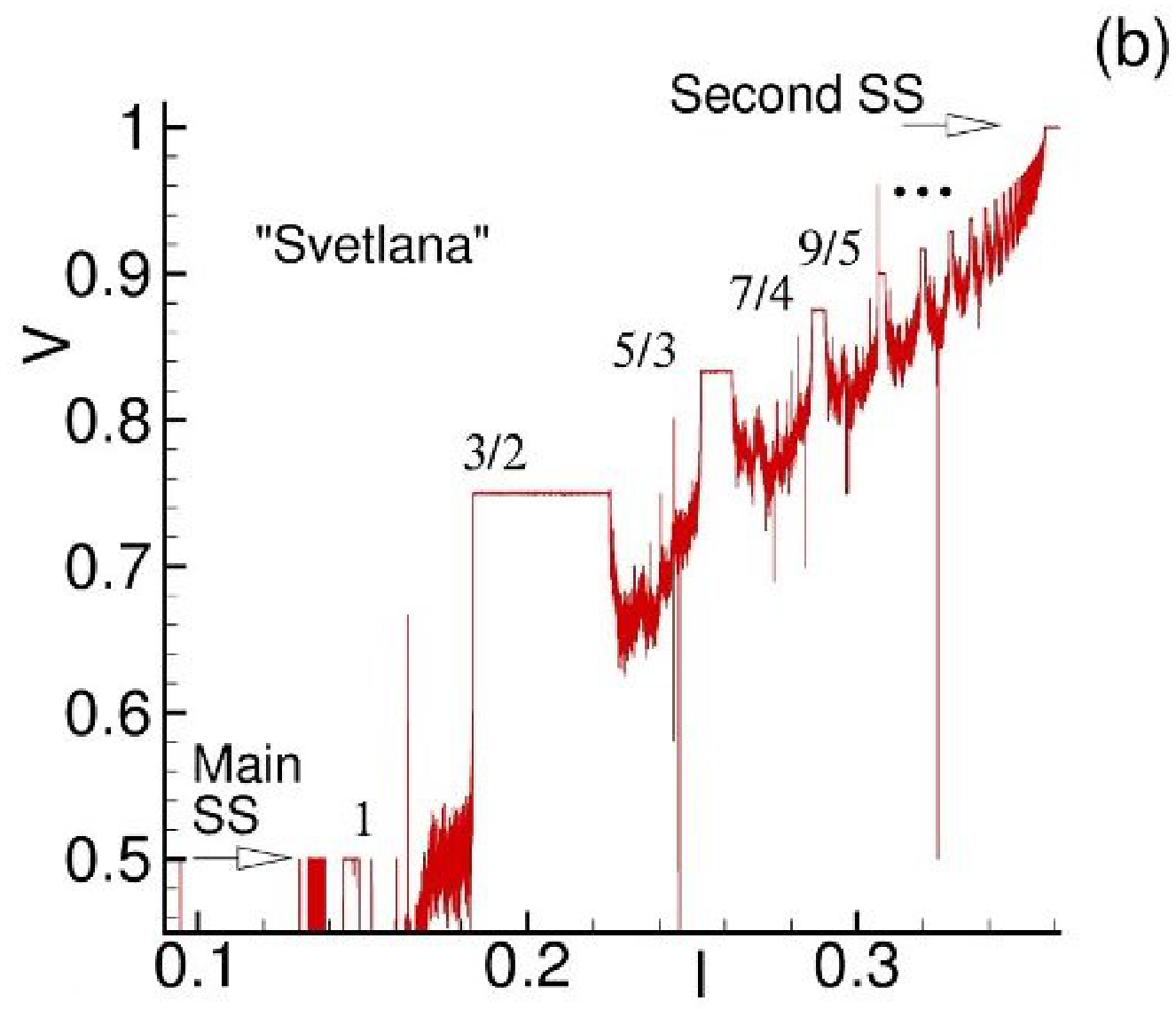}
\caption[bch!]{(Color online) (a) The IV-characteristics of the JJ at
  $\beta=0.3$, $\omega=0.5$ and $A=0.8$. Arrows with numbers indicate
  SS harmonics. The inset shows the IV-characteristic without
  radiation; (b) The enlarged part of the IV-characteristics
  (``Svetlana'') marked by  the circle in (a), demonstrating the steps
  alternating with chaotic regions. These subharmonic steps are
  numbered according to the second level of of the continued fraction
  formula $V=(2-(1/n))\omega$, with $n = 1, 2, \ldots$.}
\label{1}
\end{figure}

The IV-characteristics of the JJ with dissipation parameter $\beta<1$
are given by the Stewart-McCumber model that predicts the hysteresis
\cite{stewart68,mccumber68}. The result of a simulation of
IV-characteristics of a JJ at $\beta=0.3$, and without radiation is
presented in the inset to Fig.~\ref{1}(a).
The addition of external electromagnetic radiation, characterized by
the additional current  $A\sin({\omega t})$ in Eq.~(\ref{current}),
leads to the appearance of Shapiro steps and their subharmonics. At
high amplitude of radiation the hysteresis disappears. The
IV-characteristics of the JJ at $\omega=0.5$ and $A=0.8$ is shown in
the main part of Fig.~\ref{1}(a). We see that there is no
hysteresis in comparison with the case at $A=0$  or small $A$ (see
Ref.~\cite{S1}) and chaos develops within some current
intervals. Figure~\ref{1}(a) indicates that the model has the
propensity for both resonance and chaos. For analysis of the observed
Shapiro steps and their subharmonics, we use an algorithm proposed in
Ref.~\cite{S1}. According to it the steps in the  staircase structures
form continued fractions with voltages determined by

\begin{eqnarray}
V=\left(N\pm\frac{1}{n\pm\frac{1}{m\pm\frac{1}{p\pm
      \ldots}}}\right)\omega  \mbox{,} \label{cf}
\end{eqnarray}
where $N,n,m,p,\ldots $ are positive integers. Truncating
Eq. (\ref{cf}) at $N$ gives the first-level terms of the continued
fraction, corresponding to the main Shapiro step, or
harmonics. Similarly, truncating the formula at $n$ gives the
second-level terms, corresponding to subharmonic Shapiro steps, etc.

There is a manifestation of a Shapiro step at $V=\omega=0.5$ and its
harmonics at $V=1,1.5,2,2.5,3$, as shown in Fig.~\ref{1}(a) by
arrows below the corresponding numbers. In the chaotic region we
observe transitions between states related to the different SS
harmonics, as the current varies. In particular, very intensive
transitions occur between the second, third and fourth SS
harmonics. The IV-curves in Fig.~\ref{1} were obtained by sweeping
the bias current from $I=0$ to $I=1.2$ and back to zero. Both curves
(obtained by going up and down in current) coincide, showing no
hysteresis.

\begin{figure}[h!]
\includegraphics[height=65mm]{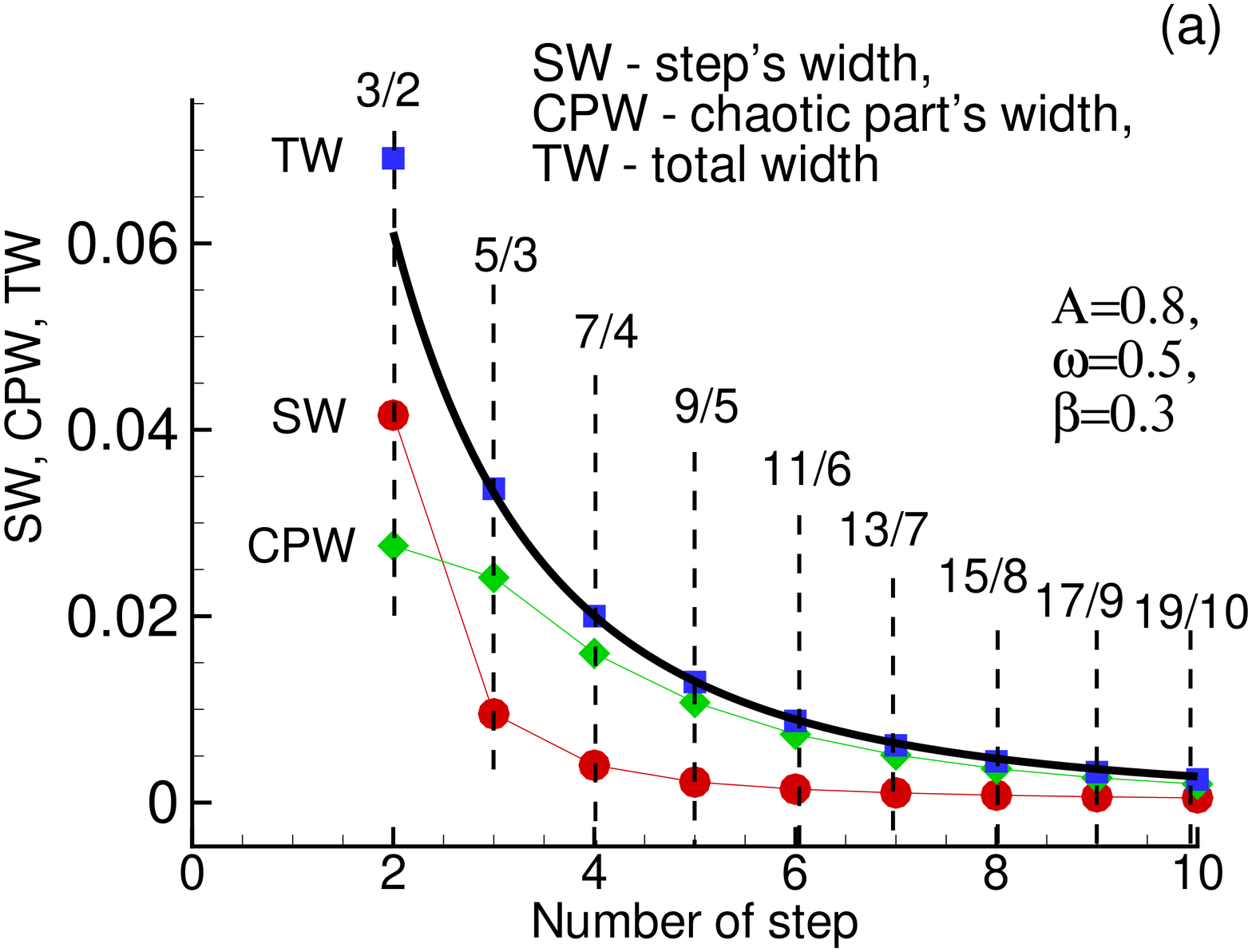}
\includegraphics[height=65mm]{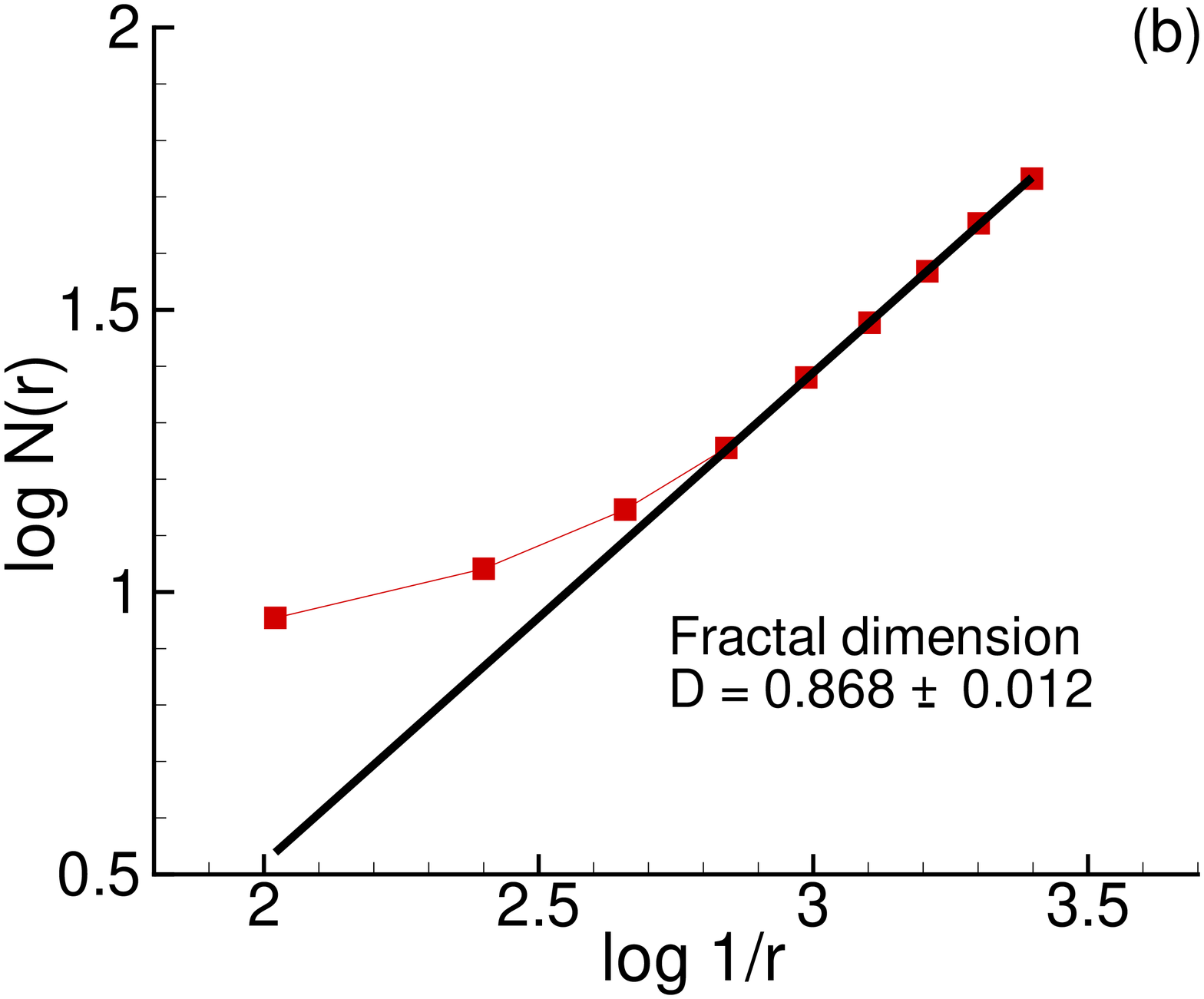}
\caption[bch!]{(Color online) (a) The width of the steps (SW,
  circles), chaotic part widths (CPW, diamonds) and total width (TW,
  squares) for the Svetlana. The total width is the sum of SW and
  CPW. The solid black line is fitted from a Bohr, hydrogen-atom-type
  equation; (b) Determination of the fractal dimension $D$ and its
  uncertainty via the box counting method \cite{hilborn00}. See main
  text for details.}
\label{2}
\end{figure}

In this paper we concentrate on that part of the IV-characteristic
marked by circle in Fig.~\ref{1}(a). An enlarged view of the
encircled region is shown in Fig.~\ref{1}(b), where a series of
steps, in the form of $(N-(1/n))\omega$, is observed between $\omega$
and $2\omega$, i.e. for $N=2$ and $n$ a positive integer. We note that
these steps approach the second Shapiro step from below. An
interesting feature of this staircase is the structured chaos
(alternating changes of the steps and chaotic regions, with changing
bias current) that materializes along with it; i.e., we have
alternating windows of resonance and chaos. We name this structure
``Svetlana'' \cite{svetlana}. The manifestation of such type of
structures are present in the numerical simulations of other authors
as well, although it has not been identified explicitly before.
Particularly, in the structure between the forth and fifth SS, shown
in the inset to Fig.13 of Ref.~\cite{noeldeke86}, we can see the
subharmonic steps separated by chaotic parts. But in the present case
the Svetlana structure represents the complete structure between two
SS (i.e. between $V=0.5$ and $V=1$). The steps appear at voltages
$\omega$, $3\omega/2$, $5\omega/3$, $7\omega/4$, etc., and approach
the second SS harmonic at $V=1$. The regular step width (SW) and
chaotic part width (CPW) of the subharmonics in Svetlana are presented
in Fig.~\ref{2}(a). We see that starting from the step $5/3$ the
width of the chaotic part to the right of each step becomes larger
relative to the step width, while both parts decrease
monotonically. The total width (TW), i.e. SW + CPW, is also shown in
the figure, in this case by the square markers.

In view of the chaos that appears in the Svetlana, it is interesting
to ask whether or not the structure is still complete, as in the case
of the devil's staircase. In mathematical terminology the staircase is
complete if the set not covered by the steps is of measure zero. For
the complete devil's staircase this corresponds to a universal fractal
dimension of close to $0.87$ \cite{jensen83}. To estimate the fractal
dimension of the Svetlana we followed the method described in
Ref.~\cite{hilborn00}. Our results are summarized in
Fig.~\ref{2}(b), where we have determined the dimension $D$ from
the slope, as $r \longrightarrow 0$, of a log-log graph of $N(r)$
against $1/r$. Here $N(r)$ is the minimum number of boxes of dimension
$r$ required to contain all the points in the geometric object. As can
be seen in Fig.~\ref{2}(b), the number of boxes increases as the
box size reduces, and in the limit as $r$ becomes smaller the slope of
the graph becomes constant. By performing a linear regression on the
linear part of the data plotted in Fig.~\ref{2}(b), i.e. using only
the six points to the right of the figure, we obtained a dimension of
$D=0.868$, to with an uncertainty of  $\pm 0.012$.

An interesting phenomenological feature that can be used to quantify
the results in Fig.~\ref{1}(b) is the trend that we observe in the
width of the steps and the chaotic intervals interleaving them. We
find that an equation of the same form as that of the Bohr atom, given
by
\begin{eqnarray}
l={l_0}\left({1/{n^2} - 1/{(n+1)^2}}\right) \mbox{,} \label{balmer}
\end{eqnarray}
describes the widths of the aforementioned intervals well. In this
equation $l_0$ is constant for the whole set of steps, and $n$ is
particular to a step. As we ascend the staircase, we go to higher $n$,
consecutively. We find that the formula works to a high accuracy,
irrespective of whether we only take the width of the steps, or we
include the width of the chaotic windows, as if the staircase were
intact. The thick line in Fig.~\ref{2}(a) is a fit of
Eq.~(\ref{balmer}) to the simulation data.  The curious fact that the
data is fit so well by Eq.~(\ref{balmer}) may be understood if we view
the Bohr equation as a Pad\'{e} approximant of an inverse cubic
polynomial. Thus it has nothing to do with quantum mechanics. Such a
fitting formula is nevertheless important for the present work because
it determines the scaling properties of our results.

\section{STRUCTURED CHAOS}
\begin{figure}[tph!]
\includegraphics[height=65mm]{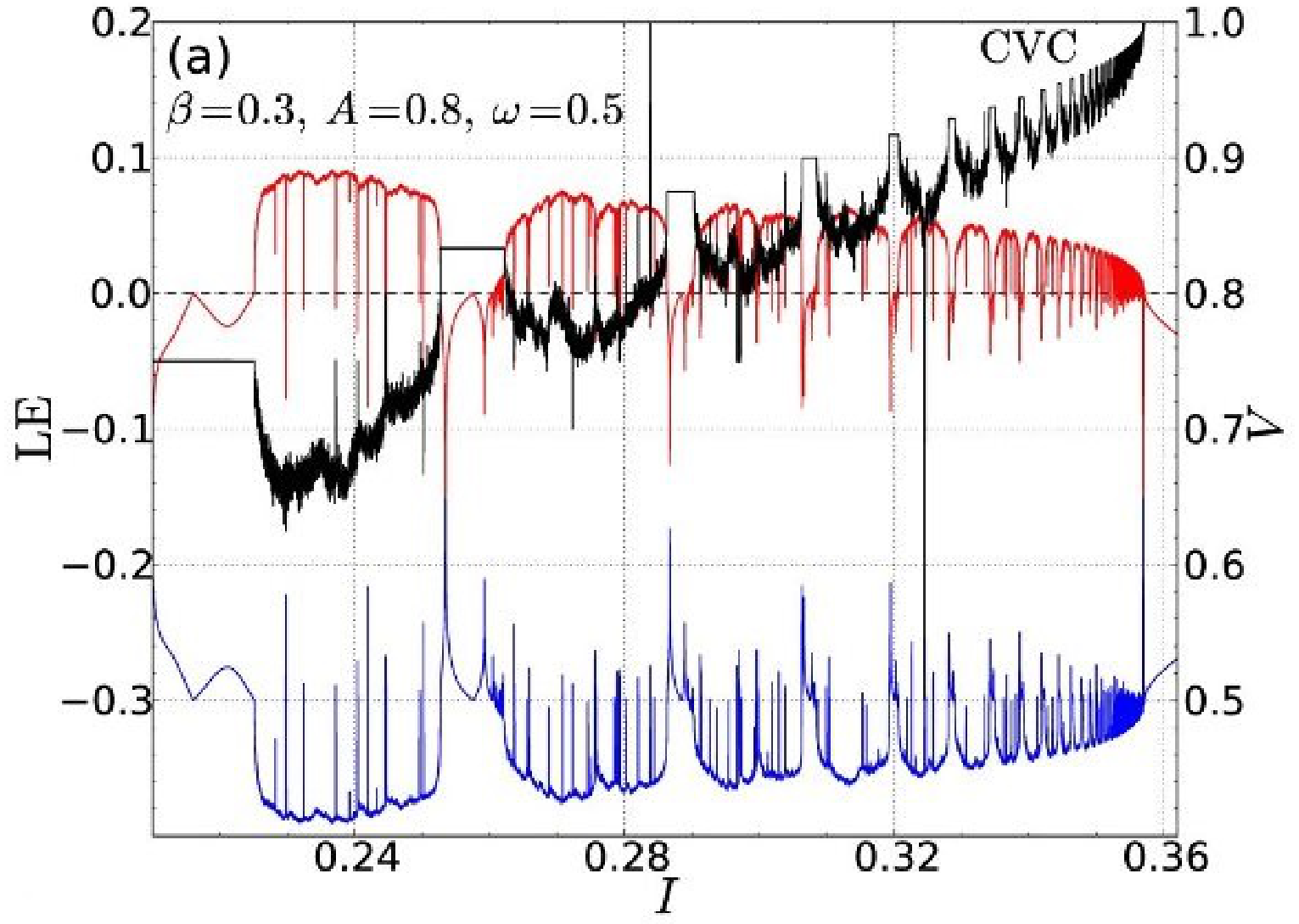}
\includegraphics[height=65mm]{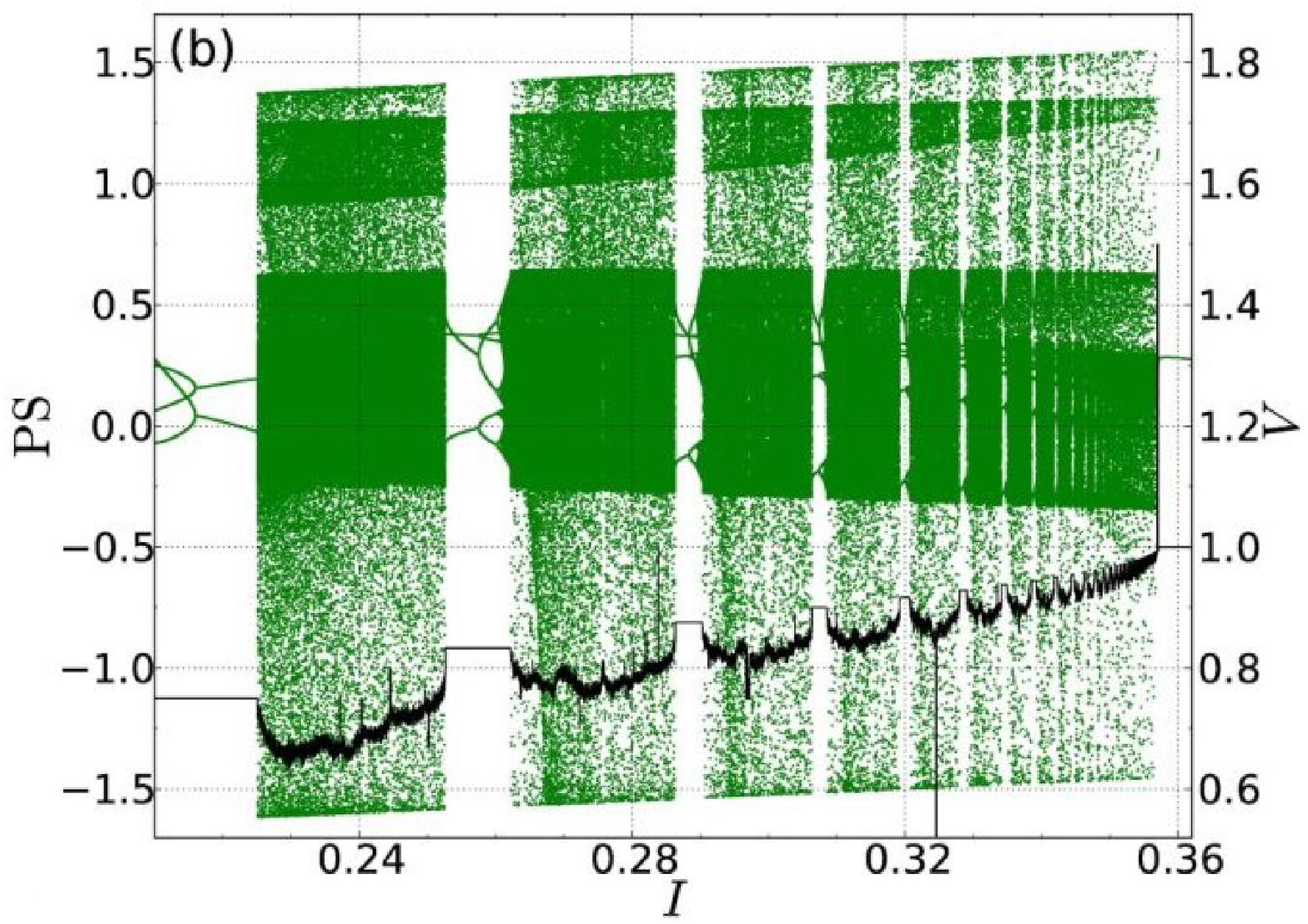}
\caption[bch!]{(Color online)  High precision calculation of (a) the
  Lyapunov exponents and IV-characteristic and (b) the Poincar\'{e}
  section and IV-characteristic (repeated for clarity) of the Svetlana
  structure.}
\label{3}
\end{figure}

Portions of the IV-characteristics between neighboring steps in
Fig.~\ref{1}(b) result from chaotic dynamics.  Such chaotic nature
is confirmed by the calculation of the Lyapunov exponents (LE) and
Poincare section (PS). Figure~\ref{3}(a) shows  the two non-trivial
LEs (thinner red and blue lines) and IV-characteristic (thicker black
line)  as functions of the dc--bias current. The two LE obey the sum
rule \cite{kautz85a} $\lambda_1 + \lambda_2 = -\beta$ and are
therefore mirror images of each other with respect to the horizontal
line at $-0.15$ on the left hand scale.  We see that the steps in the
IV-characteristic coincide with regular behavior, as indicated by the
negative regions of the maximal LE (shown in red online). On the other
hand, in between steps we see that the maximal LE is positive,
indicating chaotic behavior.

We note that PS analysis gives an effective method to investigate the
Shapiro step subharmonics in the devil's staircase, as well as the
chaotic dynamics.  In Fig.~\ref{3} this is complemented by
calculation of the LEs, indicating the position of bifurcations (when
they touch zero) and chaos (when they go positive). The PS provide a
vivid expression of synchronization between the Josephson frequency
and the external drive, especially for the subharmonics.

In Fig.~\ref{3}(b) we show the PS corresponding to the same
structure as in (a). Here the section has been made by plotting the
instantaneous steady state voltage, for each bias current, at
precisely the times when the magnitude of the (sinusoidal) external
radiation changes from negative to positive. For each bias current
1000 voltage points are plotted. We note that in the regular on-step
regions these voltage values coincide and form continuous curves with
respect to the changing bias current.

The number of PS curves within the regular regions, i.e. corresponding
to the steps in the IV-characteristic, coincide with the denominator
of the related term in the continued fraction sequence.  This feature
reflects the origin of the subharmonics which appear at voltage steps
given by $\omega_J= p\omega/q$, where $p$ and $q$ are positive
integers. These steps occur when the Josephson frequency is such that
$p$ cycles of the phase correspond precisely to $q$ cycles of the
external radiation. For example, the step at $V= 5\omega/3$
corresponds to the $3\omega_J=5\omega$.

Looking more closely at the bifurcation diagram we see that an
increase of the bias current along any particular step causes a series
of period doubling bifurcations which do not alter the constant value
of the average voltage on the step. Such on-step period doubling
bifurcations have been discussed in Ref.~\cite{kautz96}, for example.

\section{BACKBONE FEATURES AND FAREY SUM RULE}

\begin{figure*}[tph!]
\includegraphics[width=120mm]{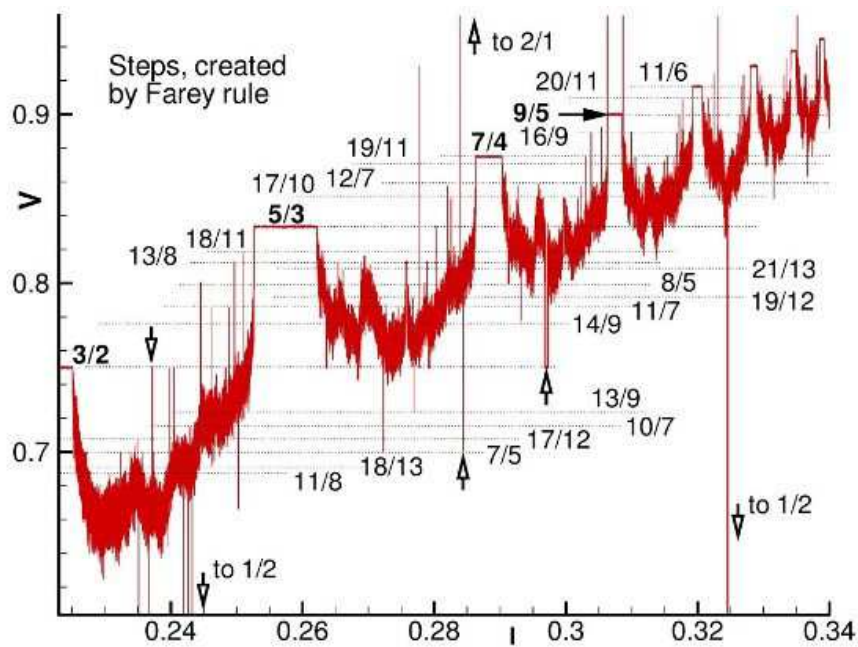}
\caption{(Color online) The underlying structure of the $IV$ staircase
  as characterized by the Farey steps, even as part of the staircase
  has been destroyed by chaos. As an example, the Farey sequence $3/2,
  11/7, 8/5, 13/8, 5/3$, is based on the Svetlana steps $3/2$ and
  $5/3$. }
\label{4}
\end{figure*}

\begin{figure}[tph!]
\includegraphics[height=65mm]{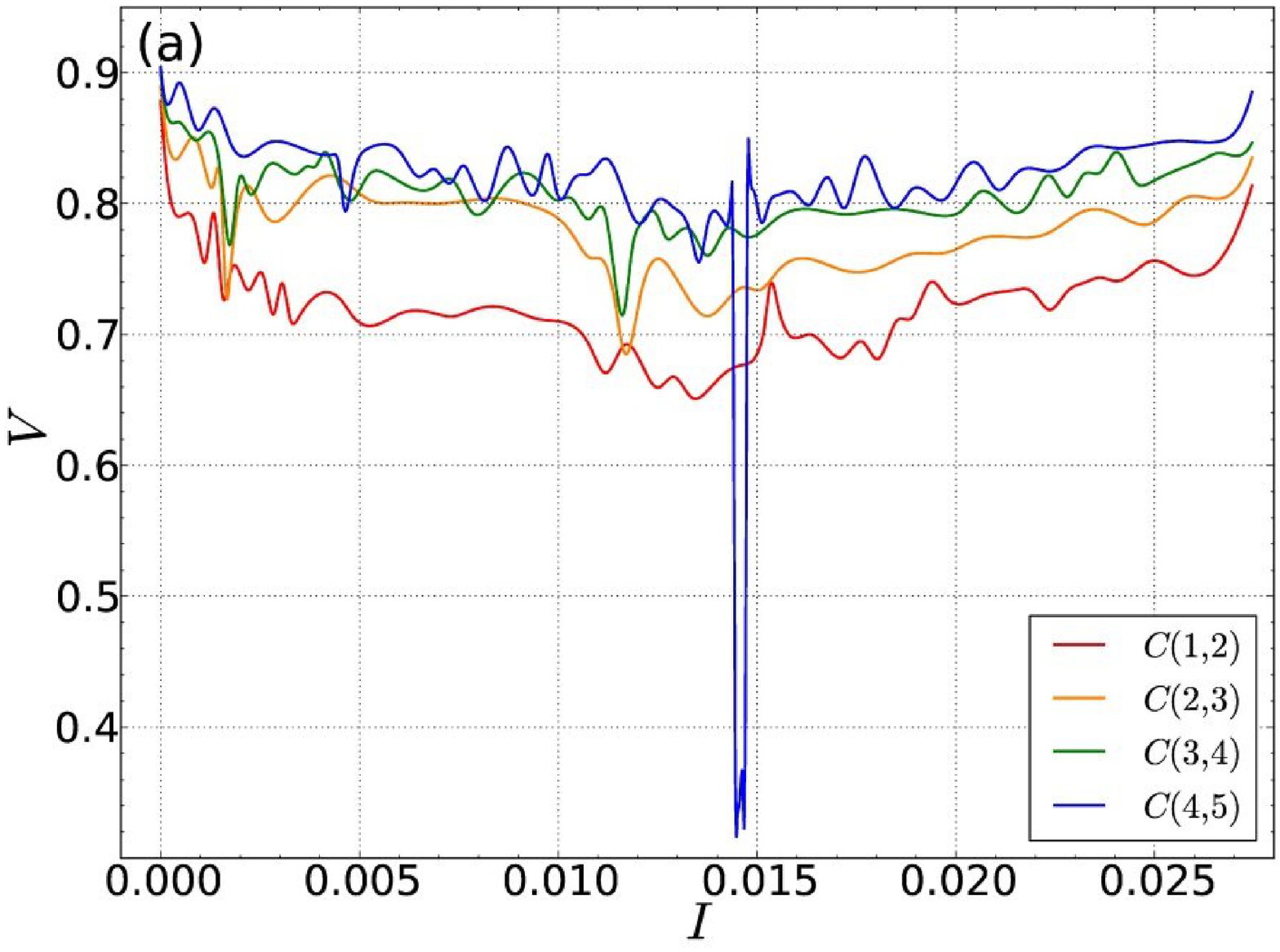}
\includegraphics[height=65mm]{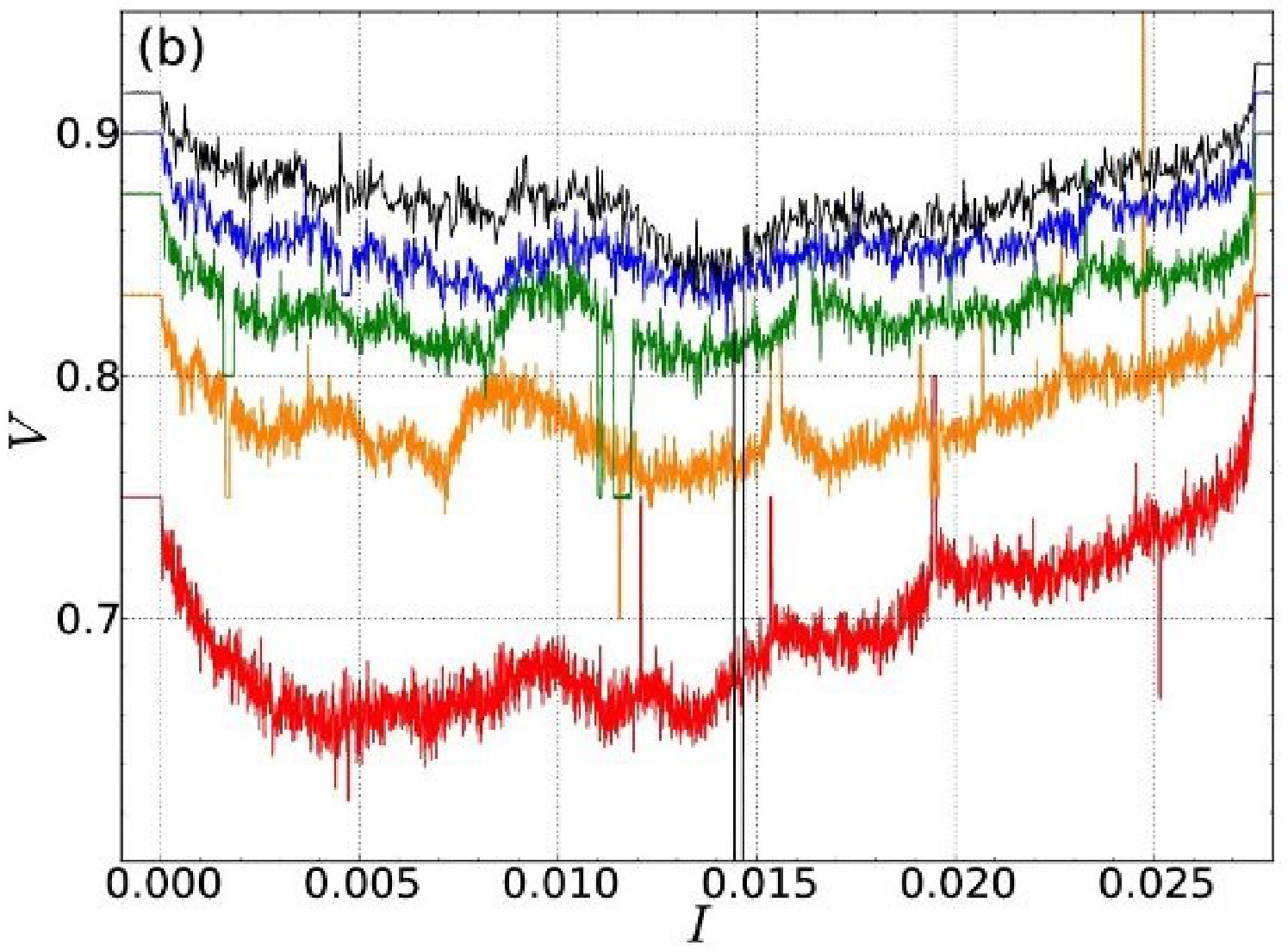}
\includegraphics[height=65mm]{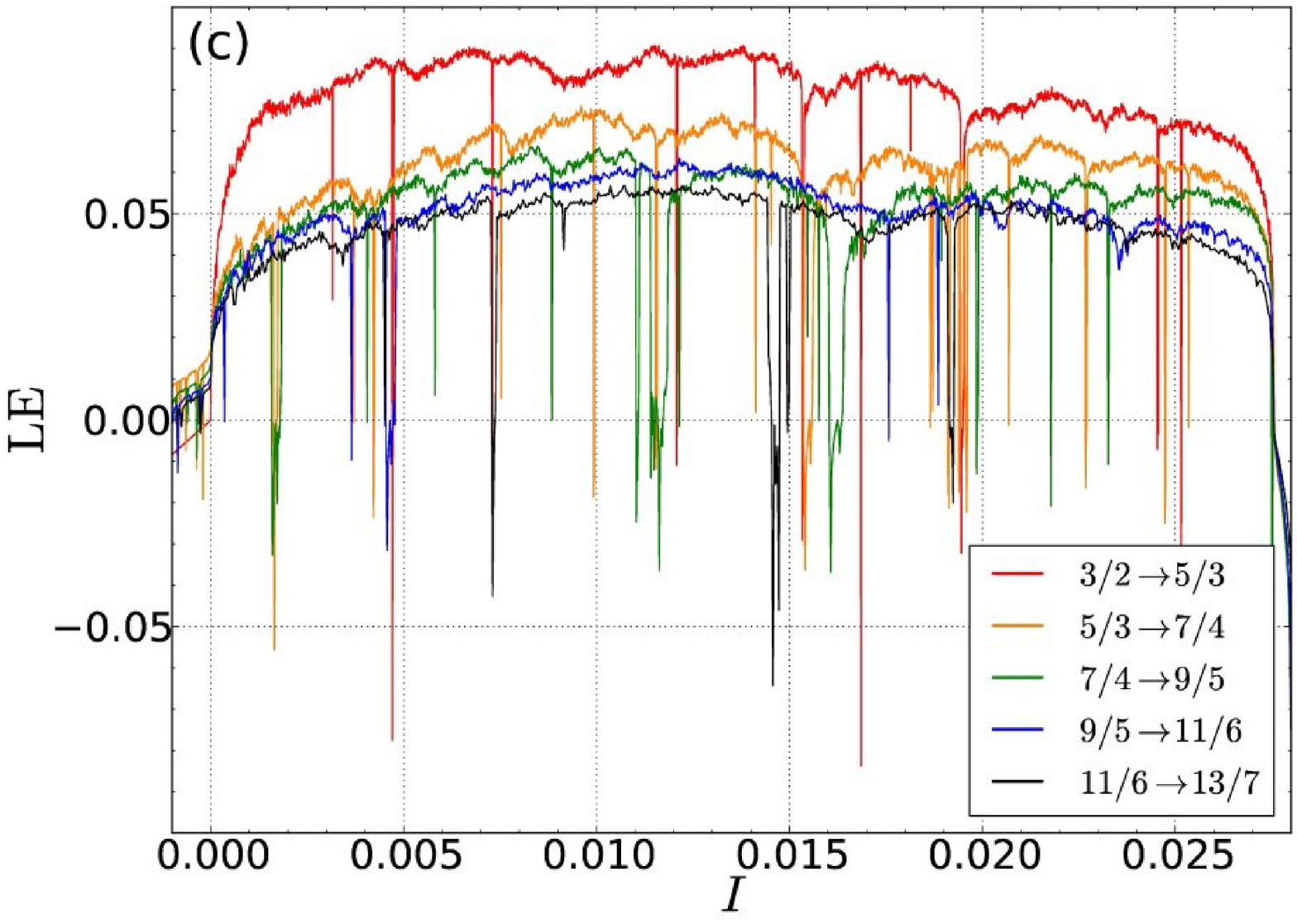}
\caption[bch!]{(Color online) The scaling that brings structure to the
  chaotic windows. This is manifested in (a) that shows the pair
  correlations (1,2), (2,3), (3,4), (4,5), each between the
  consecutive chaotic windows.  Another measure for this correlation
  is given in (b), where the average voltage is given on the scaled
  current axis; (c) The Lyapunov exponent for the dynamics at the
  scaled current.}
\label{5}
\end{figure}

The distinguishing feature of the Svetlana is structured chaos. By
structured chaos, we mean that the Svetlana can be understood as a set
of steps belonging to a devil's staircase, that has been destroyed in
a systematic way, so as to preserve the scaling properties of the
original staircase. Here, we present heuristic arguments for this
assertion, based on the results shown in Fig.~\ref{2}--Fig.~\ref{5}.

The scaling shared by the various sections of the steps, even the
irregular parts, is shown in Fig.~\ref{2}(a). The fractal property of the
staircase is shown in Fig.~\ref{2}(b). The latter property is a result of
the synchronization between the Josephson frequency and the external
radiation frequency, and it has been shown that the resulting
structure seen in the Shapiro steps can be accurately reproduced by a
continued fraction formula \cite{S1}. The different levels of this
continued fraction form a self-similar structure, called the devil's
staircase \cite{bak86}. In this sense, Svetlana is what is left of a
staircase that was to form a devil's staircase, and yet Svetlana keeps
the scaling even in the irregular windows, again as seen in
Fig.~\ref{2}. This hints at an underlying structure that these apparently
irregular regions have in common. Fig.~\ref{3} shows that these irregular
variations of voltage with current refer to the chaotic dynamics of
the system. We should note that the chaotic windows do not show
monotonic variations of average voltage with current.  This is in
contrast to the structure of a complete devil's staircase; we will
return to this aspect of Svetlana.

According to Fig.~\ref{2} and Eq. (4), scaling in current can provide
insight into the behavior of the junction, even when comparing the
chaotic regions.  We can map the chaotic windows onto each other using
a linear mapping. For the purposes of the following analysis, we
consider the first five chaotic windows that interleave the six steps:
$3/2$ (for $n=2$), $5/3$ (for $n=3$), $7/4$ (for $n=4$), $9/5$ (for
$n=5$), $11/6$ (for $n=6$), $13/7$ (for $n=7$). For simplicity we will
refer to these windows as $1$ through $5$. We can define a correlation
function as usual; i.e. by averaging over the product of voltages at
corresponding times, at a given current, and normalizing to the
self-correlated values. We have,

$$ C(V_{I_{i\alpha }},V_{I_{j\alpha
}})={\frac{{\int_{0}^{T}V_{I_{i\alpha }}(t)V_{I_{j\alpha
      }}(t)\mathrm{d}t}}{\sqrt{({\int_{0}^{T}V_{I_{i\alpha
        }}^{2}(t)\mathrm{d}t})({\int_{0}^{T}V_{I_{j\alpha
        }}^{2}(t)\mathrm{d}t})}}}.\eqno (5)
$$%
Here, $I_{i\alpha }$ denotes a specific current within the $i$th
chaotic region ($i=1,2,3,4,5$), in which each region has been divided
into $N$ equal current steps, i.e. $\alpha =1,2,\ldots ,N$.
Equivalent currents are then defined by the linear relationship
$I_{i\alpha }=a_{j}\left( I_{j\alpha }-I_{j0}\right) +I_{i0}$, where
the $a_{j}=\left( I_{iN}-I_{i0}\right) /\left( I_{jN}-I_{j0}\right) $
are scaling factors. In this work we chose $N=550$ and made use of
Simpson's $3/8$ rule to evaluate the integrals over a time domain of
$T=2000$ dimensionless units. We note that in calculating the
correlation integrals, it is important to ensure that the phase of the
driving force, i.e. the $\sin \left( \omega t\right) $ term in
Eq. (1), is the same at the start of the integration ($t=0$) for both
signals ${V_{I_{i\alpha }}(t)}$ and ${V_{I_{j\alpha }}(t)}$. If this
is not the case, the correlations $C(V_{I_{i\alpha }},V_{I_{j\alpha
}})$ are practically zero, as one might expect.

To bring out the scaling property of the step, specially the chaotic
region that is now making up a part of it, as Fig.~\ref{2}(a) is also
showing, we compare the mentioned chaotic windows $1$ through $5$, in
Fig.~\ref{5}. As just mentioned, the current axis is scaled so that the
chaotic regions for steps 1 through 5 span the same current interval
(in length) as step 1 does. In Fig.~\ref{5}(a) we have four of the voltage pair correlation functions, using Eq. (5) for these five chaotic
regions.  Fig.~\ref{5}(b) shows the voltages for the five chaotic regions on this scaled axis, and Fig.~\ref{5}(c) shows the Lyapunov exponents.

A closer look at Fig.~\ref{1}(b) with the aim of revealing its fine
structure is given by Fig.~\ref{4}. The widest phase-locked region; i.e. the
widest step, between any two resonances $p/q$ and $P/Q$ is known to be
given by $(p+P)/(q+Q)$ known as the Farey sum of the two rational
numbers \cite{hilborn00}. The Farey sum resonance is included in the
continued fraction, but at a level further down. The Svetlana steps
are used to obtain the Farey steps; e.g. the sequence $5/3, 17/10,
12/7, 19/11, 7/4$, is based on $5/3$ and $7/4$. We have shown the
position of the Farey steps on top of the Svetlana, to emphasize the
underlying structure in the blurred and chaotic regions.

In order to unify and explain these results, we highlight the presence
of a multiple devil's staircase, associated with each step
\cite{qu98}.  The results point to the equivalence of our model at the
specified parameters, to a discrete map with two discontinuities. The
usual devil's staircase is a monotonically increasing function, as
manifested in the circle map. When discontinuities are introduced,
this monotonicity is lost, and the phase-locked regions compose
tower-like structures. That is, each step forms a mesa, associated
with two devil's staircases, one ascending and leading to it; the
other descending from it, until it joins the ascending staircase
leading to the next plateau, given by the next step. This
non-monotonic behavior in voltage as a function of current, as well as
the formation of steps as mesas is clearly shown in Fig.~\ref{4}. Another
interesting feature of this figure is the brief presence of the
various steps, as emphasized by the Farey steps. This is also
confirmed by the negative Lyapunov exponents that signal remnants of
the staircase. Based on these observations we conjecture that each
chaotic interval is the result of resonance overlap in this backbone
structure.  The idea is that the chaos is the result of the
deterministically chaotic wandering of the dynamics amongst the
resulting Farey resonances, as they start to overlap. This is the main
argument leading to what we have termed as structured chaos in the
devil's staircase.

\section{Chaos on the subharmonic steps of the Svetlana}
Although it has previously been shown that the onset of chaos, on
certain \emph{harmonic} steps of a single rf-biased junction, occurs
through the Feigenbaum scenario \cite{feigenbaum79,feigenbaum80}, the
onset of chaos on \emph{subharmonic} steps such as those occurring in
the Svetlana has not been investigated previously. In the present
section we will make use of extremely high-resolution simulation data
to demonstrate that the route to chaos on the subharmonic steps of the
Svetlana also occurs via the famous Feigenbaum scenario. Furthermore,
our simulations reveal another interesting property of the Svetlana;
namely, that the subharmonic steps are not immediately destroyed as
the chaos sets in. We discuss this phenomenon in subsection $C$.

\subsection{Demonstration of Feigenbaum period doubling scenario}
In the rf-biased Josephson junction, the Feigenbaum scenario was first
identified by Huberman {\em et al.} \cite{huberman80} and later
examined by many authors.  (See, for example, Sec. 5.6 of
Ref.~\cite{kautz96} and the references therein.)

\begin{figure*}[tph!]
\includegraphics[width=\textwidth]{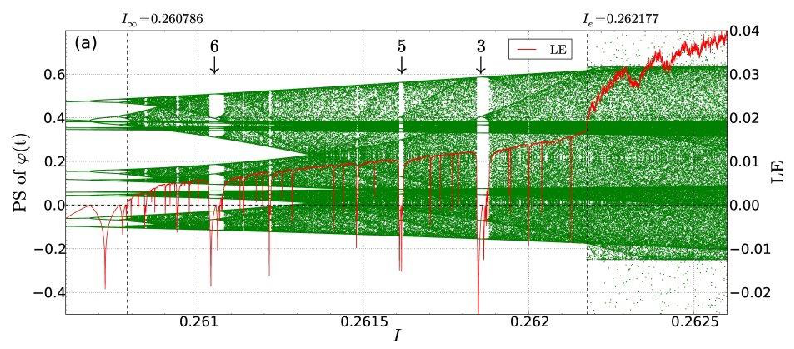}
\includegraphics[width=\textwidth]{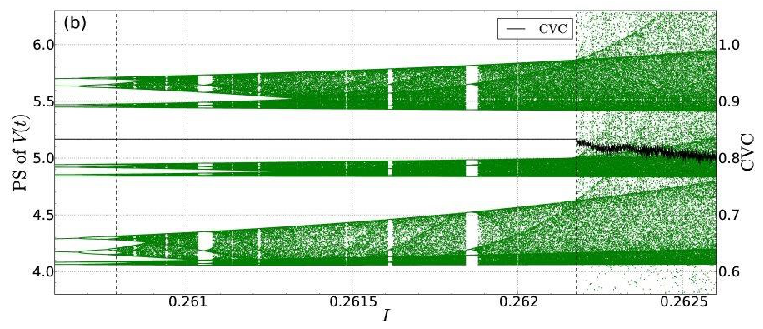}
\caption{(Color online) (a) Maximal Lyapunov exponent (solid red line)
  and Poincar\'{e} section of the phase (green dots) for bias currents
  corresponding to the right edge of the $5 \omega /3$ step; (b)
  Poincar\'{e} section of the voltage (green dots) and the average
  voltage (solid black line) over the same bias current range. In both
  figures the section plane for the PS coincides with times when the
  sinusoidal rf-bias signal crosses from negative to positive.}
\label{6}
\end{figure*}

In Fig.~\ref{6} we show the behavior of the maximal LE and PS
within a current interval corresponding to the transition from one of
the subharmonic steps in the Svetlana to the chaotic region just to
the right of the same step.

In this case we are considering the $5\omega /3$ step that was
previously shown in Figs.~\ref{1}(b) and \ref{3}. As the edge of
the step is approached with increasing bias current, the transition to
chaos occurs via a sequence of period-doubling bifurcations. This
sequence of bifurcations is infinite and, as we will show below, it
follows Feigenbaum's the well-known universal scaling laws.

To test the scaling properties for the aforementioned bifurcation
sequence we make use of the high-resolution simulation data that is
plotted in  Fig.~\ref{6}. For this simulation we used a smaller
than usual current step ($ \Delta I=10^{-8}$) to locate the first
seven values of the currents $I_{n}$ at which the successive
period-doubling bifurcation occur. The seven values so-obtained,
including the value of current at the onset of chaos $I_{\infty }$,
are listed in Table~\ref{tab1}. The values for $I_{n}$ listed in
Table~\ref{tab1}, produce successive approximations that converge to
Feigenbaum's delta, which is defined as
\begin{equation*}
\delta =\lim_{n\rightarrow \infty
}\frac{I_{n}-I_{n-1}}{I_{n+1}-I_{n}}\text{.  }\eqno (6)
\end{equation*}
These successive approximations are listed in the third column of
Table \ref {tab1}. The best approximation found is $\delta
_{6}=4.6713$, which is very close to the theoretical value of $\delta
=4.6692$.

\begin{table}[tbph]
\caption{Calculation of the Feigenbaum numbers $\protect\alpha $ and $
  \protect\delta $ for the onset of chaos on the $5\protect\omega /3$
  subharmonic step of the Josephson junction. Entries with three
  asterisks (***) could not be determined from the simulation data to
  the same uncertainty as the listed entries.}
\begin{tabular}{|c||c|c|c|c|}
\hline $n$ & $I_{n}$ & $\delta
_{n}=\frac{I_{n}-I_{n-1}}{I_{n+1}-I_{n}}$ & $d_{n}$ & $\alpha
_{n}=\frac{d_{n}}{d_{n+1}}$ \\ \hline\hline \multicolumn{1}{|l||}{$1$}
& \multicolumn{1}{|l|}{$0.25756981\;$} & *** & $ 0.20707$ & $2.956$
\\ \hline \multicolumn{1}{|l||}{$2$} &
\multicolumn{1}{|l|}{$0.26025001$} & $6.3363$ & $0.07004$ & $2.505$
\\ \hline \multicolumn{1}{|l||}{$3$} &
\multicolumn{1}{|l|}{$0.26067300$} & $4.7500$ & $0.02796$ & $2.503$
\\ \hline \multicolumn{1}{|l||}{$4$} &
\multicolumn{1}{|l|}{$0.26076205$} & $4.6943$ & $0.01117$ & ***
\\ \hline \multicolumn{1}{|l||}{$5$} &
\multicolumn{1}{|l|}{$0.26078102$} & $4.6840$ & *** & *** \\ \hline
\multicolumn{1}{|l||}{$6$} & \multicolumn{1}{|l|}{$0.26078507$} &
$4.6713$ & *** & *** \\ \hline \multicolumn{1}{|l||}{$7$} &
\multicolumn{1}{|l|}{$0.26078594$} & *** & *** & *** \\ \hline $\vdots
$ & $\vdots $ & $\vdots $ & $\vdots $ & $\vdots $ \\ \hline
\multicolumn{1}{|l||}{$\infty $} & \multicolumn{1}{|l|}{$0.26078606$}
& *** & *** & *** \\ \hline
\end{tabular}
\label{tab1}
\end{table}

By measuring the maximal voltage differences $d_{n}$, separating the
bifurcated branches shown in the PS, we also obtained three
increasingly accurate estimates of Feigenbaum's scaling parameter
$\alpha $, defined as
\begin{equation*}
\alpha =\lim_{n\rightarrow \infty }\frac{d_{n}}{d_{n+1}}.\eqno (7)
\end{equation*}
Successive values of $d_{n}$ were determined to $5$ decimal places, as
listed in the fourth column of Table~\ref{tab1}. At this level of
uncertainty the values of $\alpha _{n}$ could be determined accurately
up to $n=3$, and the estimate so obtained is $\alpha _{3}=2.503$,
which is also very close to the universal value of $\alpha
=2.5029$. Thus we have verified that the onset of chaos on the
subharmonic steps of the Svetlana occurs through Feigenbaum's famous
period-doubling scenario.

\subsection{Universality in the sequence of periodic windows}
Together with the aforementioned scaling laws for $\delta $ and
$\alpha $, another universal feature of the period doubling route to
chaos is the sequence of periodic windows that occurs once the chaos
has set in. In Fig.~\ref{6}, and in Table~\ref{tab1}, we see that
the accumulation point for the bias currents in period doubling
sequence occurs at $I_{\infty }=0.260786$. For currents higher than
$I_{\infty }$ the system is chaotic, except within a finite number of
so-called periodic windows,  which can be observed clearly in both the
PS, and the LE (the maximal LE becomes negative within each periodic
window). We have found it convenient and more accurate to locate the
periodic windows by looking for the regions over which the maximal LE
drops below zero. In the present case, this method has enabled us to
locate all the periodic windows wider than $10^{-6}$ bias current
units.  For each window, listed in the order of increasing bias
current, Table~\ref{tab2} shows the value of the bias current at the
start (column 2) and end (column 3) of the window, together with the
associated period $T$ (column 4). Note that, because there are at
least three branches in the PS for currents corresponding to the
regular part of this $5\omega /3$ step (c.f.  Figs.~\ref{3} and
\ref{6}(a)) we have normalized $T$ to three  times the external
radiation period $\tau =2\pi /\omega $.

\begin{table}[tbph]
\caption{The sequence of periodic windows occurring in the chaotic
  region shown in Fig.~\ref{6}.}
\begin{tabular}{|c||c|c|c|}
\hline & $I_{\mathrm{start}}$ & $I_{\mathrm{end}}$ & $T/(3\tau )$
\\ \hline\hline \multicolumn{1}{|l||}{$1$} &
\multicolumn{1}{|l|}{$0.260797$} & $0.260799$ & $24$ \\ \hline
\multicolumn{1}{|l||}{$2$} & \multicolumn{1}{|l|}{$0.260838$} &
$0.260841$ & $12$ \\ \hline \multicolumn{1}{|l||}{$3$} &
\multicolumn{1}{|l|}{$0.260936$} & $0.260938$ & $10$ \\ \hline
\multicolumn{1}{|l||}{$4$} & \multicolumn{1}{|l|}{$0.261033$} &
$0.261054$ & $6$ \\ \hline \multicolumn{1}{|l||}{$5$} &
\multicolumn{1}{|l|}{$0.261136$} & $0.261138$ & $10$ \\ \hline
\multicolumn{1}{|l||}{$6$} & \multicolumn{1}{|l|}{$0.261215$} &
$0.261217$ & $8$ \\ \hline \multicolumn{1}{|l||}{$7$} &
\multicolumn{1}{|l|}{$0.261479$} & $0.261481$ & $7$ \\ \hline
\multicolumn{1}{|l||}{$8$} & \multicolumn{1}{|l|}{$0.261608$} &
$0.261614$ & $5$ \\ \hline $9$ & $0.261843$ & $0.261860$ & $3$
\\ \hline
\end{tabular}
\label{tab2}
\end{table}

With this normalization the sequence of the larger periodic windows is
$6$, $  5$ and $3$, as marked by the downward pointing arrows in
Fig. \ref{6}(a), in the order of increasing bias current. This
sequence has also been reported in Refs. \cite{hilborn00,gitterman10},
albeit at different parameter values.  Because this sequence of the
main periodic windows has also been observed experimentally in a wide
variety of other, quite different dynamical systems, it has become
known as a universal sequence (U-sequence). Note that in
Table~\ref{tab1} we have not listed the periods of the period-doubling
cascades which break up the larger periodic windows that have been
marked in Fig.~\ref{6}(a).  However, closer examination of these
windows reveals that their breakup also occurs through an infinite
sequence of period doubling bifurcations, which is self-similar to the
original sequence described above, in the sense that it not only obeys
the Feigenbaum scaling laws, but again contains the universal sequence
($6$, $5$ and $3$ etc.) of periodic windows at a much finer scale.
These tiny periodic windows are of course virtually impossible to
observe experimentally. Our simulation data, however, clearly predicts
that the chaos observed on the subharmonic steps of the Svetlana, also
contains a U-sequence of stable periodic orbits.

\subsection{On-step positive Lyapunov exponent}
An interesting feature of the chaotic region shown  between
$I_{\infty}$ and $I_{e}$ in Fig. \ref{6}, is the on-step positive
Lyapunov exponent. By this we are referring to the fact that the
average voltage throughout this range of bias current remains
constant, even though the LE becomes positive, i.e. the system becomes
chaotic. This type of behavior has also been observed in the numerical
simulations of Kautz \cite{kautz96}, where it was reported for a
transition to chaos that occurs on a harmonic step. In the present
case we observe the same behavior on a subharmonic.

The persistence of the average voltage step, even when the system
trajectory becomes chaotic, can be understood by looking at the PS of
the initial phases shown in Fig. \ref{6}(a). Between $I_{\infty }$
and $I_{e}$ it is clear that the initial phases remain range bound,
approximately between $-0.19$ and $0.61$. On the other hand, for bias
currents larger than and $I_{e}$ the phases span the full range
$(-\pi,\pi]$. For this reason, the chaos that occurs  between
  $I_{\infty }$ and $I_{e}$ was referred to by Kautz \cite{kautz96} as
  phase-locked chaos. However, unlike the case of an harmonic step,
  where the phase always advances by approximately $2\pi$ on every
  rf-cycle; in the present case the rf-amplitude is sufficiently large
  to cause reversals, so that in the present case the phases may
  either advance or reduce by approximately $2\pi$ on every
  rf-cycle. If one interprets this behavior in terms of a particle on
  a washboard potential, then the particle is able to advance or
  retreat by one or two wells on any given rf-cycle, which means that
  it is diffusive in nature, in spite of being phase-locked. (Note
  that, as defined by Kautz, the term phased-locked refers to the fact
  that phase is highly correlated (but not strictly-speaking locked)
  to the rf-cycle.) Thus, the distinction made between phase-locked
  and diffusive chaos in Ref.~\cite{kautz96}, breaks down in the
  present case and more work would be required to investigate the
  transition that occurs here, in going from the phased locked region
  between $I_{\infty }$ and $I_{e}$, to the diffusive region, for
  currents above $I_{e}$. For the case of an harmonic step without
  reversals, the transition occurred via a chaotic
  crisis.\cite{kautz96}

In the present article we will not investigate the nature of this
transition in greater detail. It suffices to note that, as the current
is increased beyond $I_{e}=0.262177$, the $5\omega /3$ step breaks up
as the average voltage becomes erratic. It can be seen in
Fig.~\ref{6}(a), that the disintegration of the step is accompanied
by a sharp increase in the already positive LE.

\section{DEPENDENCE OF SVETLANA ON RADIATION AND JJ PARAMETERS}
At a fixed frequency $\omega=0.5$ and dissipation parameter
$\beta=0.3$ the most pronounced structured chaotic formation
(Svetlana) is observed around $A=0.8$. A natural question appears: can
changes of JJ or radiation parameters lead to a devil's staircase
structure without chaotic windows? That is, does Svetlana appear as an
intermediate structure of synchronized steps before transition to
chaos, or does what we have come to term  structured chaos ``possess
equal rights'' in JJ dynamics under radiation, as for other types of
devil's staircase formations?

\begin{figure}[tph!]
\includegraphics[height=65mm]{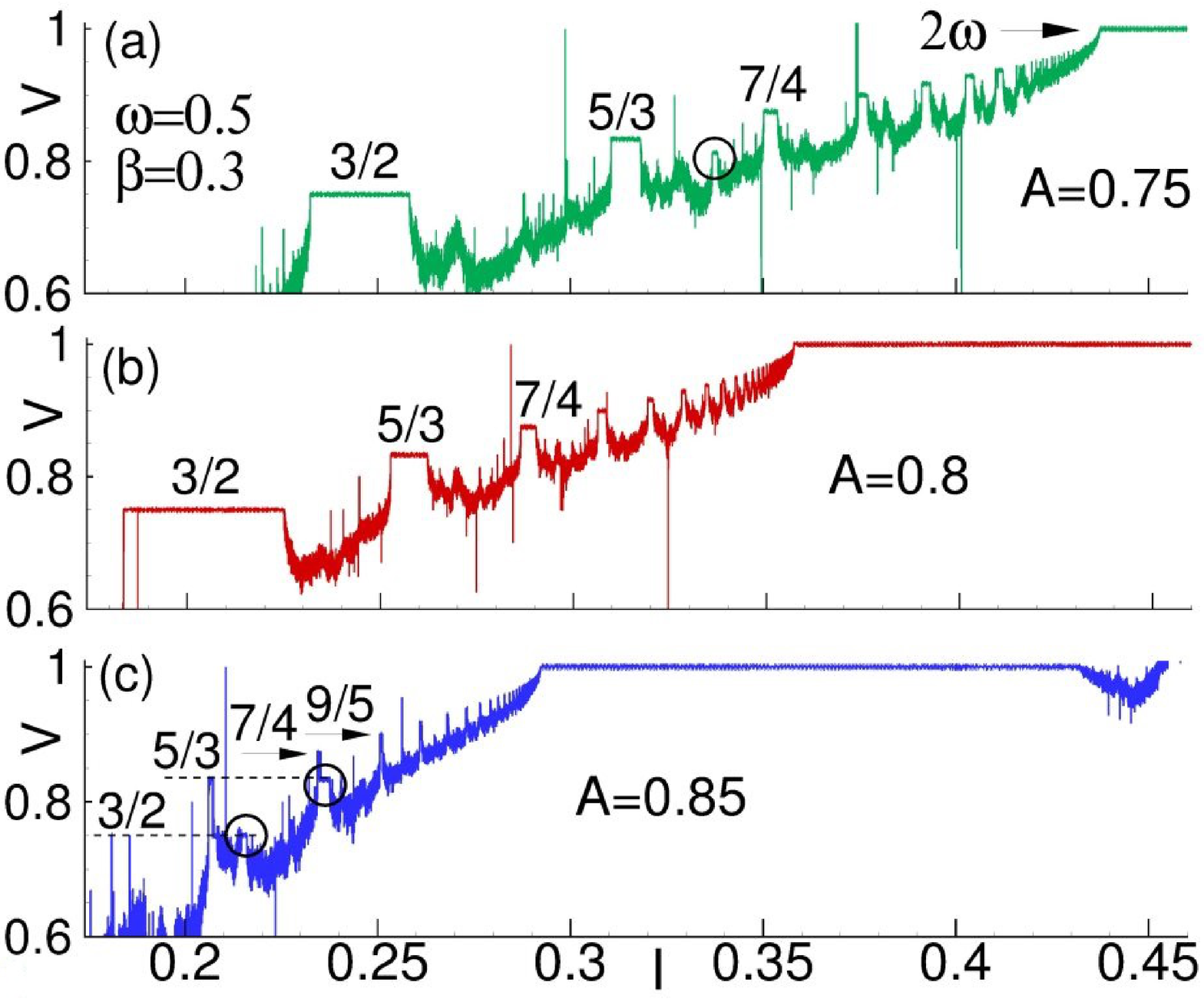}
\includegraphics[height=23mm]{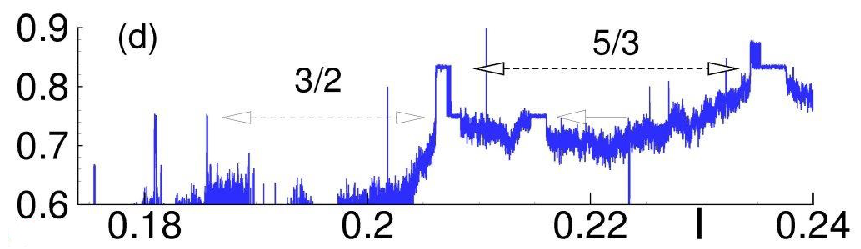}
\caption[bch!]{(Color online) A-variation of the devil's staircase
  structure Svetlana at $\omega=0.5$, $\beta=0.3$. (a) $A=0.75$; (b)
  $A=0.8$; (c) $A=0.85$; (d) Enlarged part of structure at $A=0.85$,
  demonstrating overlapping of subharmonics.}
\label{7}
\end{figure}

To answer this question we discuss here the variation of the Svetlana
with changes in the amplitude of radiation. The structures  at  three
values of the amplitude, $A=0.75, 0.8, 0.85$ are presented in
Fig.~\ref{7}. As we have seen in Fig.~\ref{1}(b), the main SS
harmonics at $V=0.5$ is not stable at the chosen JJ and radiation
parameters, so we analyze the features between the subharmonic
$3\omega/2$ and the second SS at $V=2\omega$. The known fact is that,
with an increase in amplitude of external radiation, the width of
Shapiro harmonics changes, following the Bessel function dependence
\cite{barone82,kleiner04}. Analysis shows (and it is clear from
Fig.~\ref{7}) that with an increase in $A$ the position of the
onset of the second SS harmonic at $V=2\omega$ is shifting more in
comparison to the shifting of the first SS harmonic at
$V=\omega$. That is, the current interval between the beginning of the
second SS harmonics and the end of the first one is reduced at $A=0.8$
in comparison to $A=0.75$ and increased in comparison to
$A=0.85$. These changes are consistent with the expected Bessel
function dependence.

First, we consider the effect of decreasing the radiation amplitude by
comparing the two structures at $A=0.75$ and $A=0.80$, as shown in
Figs.~\ref{7}(a) and (b). We see that the decrease in $A$, from
$A=0.80$ to $A=0.75$,  shifts the devil's staircase structure to the
right side of the considered current interval. The overall structure
is stretched,  while the widths of the steps are decreased, i.e. the
widths of the  chaotic windows increased as $A$ decreased from
$A=0.80$ to $A=0.75$.  This trend continues with a further decrease in
the amplitude of radiation (not shown here), eventually leading to the
disappearance of the Svetlana structure. Thus we do not observe the
structure without chaotic windows. As $A$ is reduced, we note the
appearance of an additional step between $5\omega/3$ and $7\omega/4$,
marked by a circle in Fig.~\ref{7}(a). This additional step belongs
to the third continued fraction level, determined by
$V=\left(N\pm\frac{1}{n\pm\frac{1}{m}}\right)\omega$ with $N=2$,
$n=3$, and $m=3$. It is the $13\omega/8$ step, with the corresponding
value of voltage being $0.8125$. This finding is in agreement with our
hypothesis about the role of the backbone in the Svetlana (see Chapter
IV).

Second, we consider the effect of increasing the amplitude of
radiation by comparing the structures at $A=0.8$ and $A=0.85$ as
presented in Fig.~\ref{7}(b,c). There is a shift of the devil's
staircase structure to the left side of considered current interval,
and then there is a decrease in the width of the SS subharmonics.  The
current interval between the beginning of the second SS harmonics and
the end of the first one is reduced at $A=0.85$ in comparison to
$A=0.8$. This shrinking of the devil's staircase structure has the
same explanation as the stretching at $A=0.75$. We note also the
appearance of the $3\omega/2$-subharmonics in the current interval
between $5\omega/3$ and $7\omega/4$ subharmonics in Fig.~\ref{7}(c)
and the appearance of the $5\omega/3$ subharmonics in the current
interval between $7\omega/4$ and $9\omega/5$ subharmonics. The
corresponding part of IV-characteristic is enlarged in
Fig.~\ref{7}(d). These facts demonstrate an overlapping mechanism
of the transition of synchronized steps to the chaotic state. With
further increase in $A$ the chaotic regions are expanded and devil's
staircase structure disappears. So, an increase in the amplitude of
radiation leads to a decrease in the subharmonic widths and a relative
increase in the chaotic regions.

\begin{figure}[tph!]
\includegraphics[height=65mm]{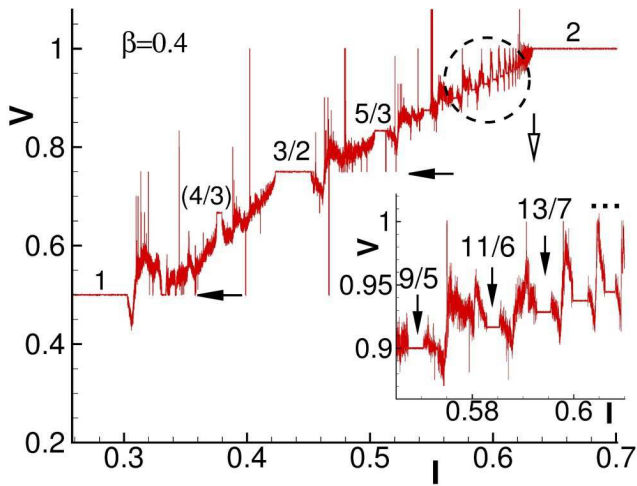}
\caption[bch!]{(Color online) The devil's staircase structure at
  $\omega=0.5$, $A=0.8$ and  $\beta=0.4$. Inset enlarges the part of
  the structure marked by circle.}
\label{8}
\end{figure}

The chaotic parts of Svetlana can be essentially modified by changing
of JJ parameters. Figure~\ref{8} presents the devil's staircase
structure in Svetlana at the same radiation parameters $\omega=0.5$
and $A=0.8$, but at a different dissipation parameter $\beta=0.4$. We
see that the character of chaotic regions is changed, but it still
demonstrates regular variation of the steps and chaos. We have also
observed here the additional step $4/3$ (taken in the brackets in
figure) which belongs to another continued fraction. The inset
enlarges the part of the structure marked by circle. We see clearly
the similar character of the chaotic parts for higher steps.

Thus, the answer to the question stated at the beginning of this
section is positive: based on our results, we note that the structured
chaos is a stable formation over definite intervals of $A$. Analogous
results are obtained by changing the frequency of radiation $\omega$.

\section{Analysis of the Experimental Results}

Let us now briefly discuss the existing experimental results. The
experimental survey of chaos in the Josephson effect was presented in
Ref.~\cite{noeldeke86}. The authors analyzed the range of chaotic
behavior in tin tunnel junctions and indium microbridges subjected to
dc and rf bias. An important fact found by this detailed analysis was
a statement that the experimental results agree with simulations based
on the resistively and capacitively shunted junction (RCSJ) model and
theoretical predictions. This fact stresses the necessity for further
numerical investigations within the RCSJ model.

One of the interesting features of the IV-characteristics under
external electromagnetic radiation is a fragmentation of Shapiro steps
and their subharmonics. Such fragmentation was demonstrated in the
numerical results of Kautz and Monako \cite{kautz85a}. Experimental
evidence of the fragmented IV-characteristics, particularly, strongly
distorted steps and ``wiggles," which are preceded by subharmonic
voltage steps for lower rf powers in Pb/ox/PbIn tunnel junctions and
in Pb microbridges irradiated by 70-GHz microwaves were reported in
Ref.~\cite{okuyama81}.

The results of the experimental observation of a devil's staircase on
a microwave irradiated Josephson junction prepared from the bulk
polycrystalline magnetic superconductor GdMo$_{6}$Se$_{8}$ were
presented in Ref.~\cite{kuznik93}. Measurements were made at a
temperature of $1.3 \; \mbox{K}$, under  external radiation at $9.5 \;
\mbox{GHz}$. The authors considered that the possible reason for
the appearance of the DS was an overlapping of the steps. We note that the
detailed experimental investigation of the fragmented phases at
different conditions and parameters of JJ is still lacking.  The
fragmentation of SS  was also observed in our simulations at amplitude
$A<0.8$.  An increase in amplitude leads to the overlapping of the SS
and destroys them. It can be seen in Fig.~\ref{1} for SS with
$n<4$. Results will be discussed in detail, elsewhere.

Noise rise and negative differential resistance regions in a
self-resonant circuit consisting of an inductively and resistively
shunted Nb/ox/PbIn tunnel junction were observed by Miracky {\em et
  al.} \cite{miracky83}. Gubankov {\em et al.}
\cite{gubankov83,gubankov84} investigated the frequency range of
chaotic behavior in Nb/ox/PbBi tunnel junctions. Their results are in
agreement with the simulations based on RCSJ model.\cite{noeldeke86}
The distorted IV-characteristics and a strong noise rise in the
distorted regions, which was ascribed to chaotic noise were found in
Ref.~\cite{octavio84}. Reports on measurements of dc electron
transport and microwave dynamics of thin film hybrid
Nb/Au/CaSrCuO/YBaCuO planar Josephson junctions were presented in
Ref.~\cite{constantinian10}. The authors observed tunnel-like
behavior, and oscillations in sync with the applied radiation at
integer and half-integer steps. For a junction fabricated with a
\emph{c-}oriented yttrium barium copper oxide (YBCO) film,  the
devil's staircase structure was observed under microwave  irradiation
at $4.26$~GHz.

In Refs.\cite{cronemeyer85,vanneste85,iansiti85} the authors provide
further confirmation for the existence of chaotic regions in the
intermediate- and high-McCumber parameter regime. In experiments with
short Indium microbridges the devil's staircase transition to chaos
was observed. Such data can also be evaluated within the framework of
the RCSJ model \cite{noeldeke86}.

Based on the analysis of the experimental investigations of the chaos
in the JJ, we note that at present, the experimental observation of
structured chaos in the IV-characteristics under external
electromagnetic radiation (Svetlana structure) is absent. Since our
results on numerical simulations of Svetlana were obtained by the RCSJ
model, it would be very interesting to test the formation of the
Svetlana structure and its features experimentally.

\section{CONCLUSION}
We have performed high-precision numerical simulations of the
Josephson junction under external microwave radiation, within the
well-established  RCSJ (Stewart-McCumber) model. Our simulations have
indicated the presence of an unusual structure, that occurs within the
IV-characteristics of the junction. By analyzing the IV-characteristics,  Lyapunov exponents and
Poincar\'{e} sections we have identified a region of experimentally
realizable parameters for which the subharmonic Shapiro steps are
separated by structured chaotic windows. We have called this formation,
Svetlana.

Interestingly, the dynamics of the system, which follows a
deterministic path and exhibits synchronization with the external
drive in the form of constant voltage subharmonic steps (Shapiro
steps), can be disrupted while a trace of the underlying staircase
remains. Since the low rationals in the subharmonic set are more
stable and live longer, we obtain a well structured chaos as the steps
become interleaved with chaotic windows. There is thus a gradual
disappearance of order, as well as resonance overlap. The argument
implicitly requires a subharmonic region of the staircase.

An alternative way of looking at this phenomenon is in terms of
equivalent lower-dimensional chaotic maps that are interrupted  by the
ordered dynamics. Such ideas have been conjectured and explored
previously \cite{qu98}. Here we  have shown that the multiple devil's
staircase property is effectively reproduced by a  continuous, differentiable
map. We have furthermore identified the diffusive chaotic mode as
playing a role similar to the discontinuity in the map
\cite{bauer92}. In fact, the diffusive chaos contains an indication of
such a discontinuity.

Analysis of the experimental results on chaos in JJ under external
radiation have indicated that a whole series of important results
obtained by numerical calculations is still open for experimental
investigations. If experimentally confirmed, the results obtained in
the present paper could, for example, be applied to topological
superconductivity, a field which is currently being investigated
intensively \cite{lutchyn10,alicea11}. They support Majorana fermions
which are expected to be used for realization of quantum gates that
are topologically protected from local sources of decoherence.  The authors of \cite{rokhinson12}  report the
observation of the fractional a.c. Josephson effect in a
semiconductor-superconductor nanowire junction as a signature of
Majorana quasiparticles. The use of subharmonics for the detection of
the Majorana fermions is a very interesting but unsolved problem. Its
solution may provide additional information on the Majorana physics
and  may warrant special consideration in a more detailed
investigation.

Finally, we have also demonstrated universality within the diffusively
chaotic windows. Such universality is an underlying feature of the
chaotic windows; which, while being contained within the staircase
itself, are effectively being 'washed' out. We find this aspect of the
work most interesting because  it shows some sort of predictability,
even though we have chaotic dynamics: as if the ghost of the staircase
lives on!

\begin{acknowledgments}

Yu. M. S. thanks  I. Rahmonov, M. Yu. Kupriyanov, K.Y. Constantinian,
G. A. Ovsyannikov, V. P. Koshelets  for helpful discussions and
D. V. Kamanin and the JINR-SA agreement for the support of this
work. He also appreciates kind hospitality of Prof. Y. Takayama and
Prof. N. Suzuki from Utsunomiya university where part of this work was
done. Yu. M. S. and M. R. K. wish to thank the Physics Department at
the University of South Africa for a pleasant stay.

\end{acknowledgments}


\begin{thebibliography}{99}
\bibitem{svetlana} This is the first name of the third author, who first found it in her numerical simulations.
\end{thebibliography}
\end{document}